\documentclass[prl,twocolumn]{revtex4}
\usepackage{amsfonts}
\usepackage{amssymb}
\usepackage{amsmath}
\usepackage[all]{xy}

\newcommand{\ba}{\begin{eqnarray}}
\newcommand{\ea}{\end{eqnarray}}
\newcommand{\be}{\begin{equation}}
\newcommand{\ee}{\end{equation}}

\newcommand{\ed}{\end{document}}
\newcommand{\nn}{\nonumber\\}
\newcommand{\fr}{\frac}
\newcommand{\wt}{\widetilde}

\begin{document}

\title{ON DUALITY SYMMETRY IN CHARGED P-FORM THEORIES}
\author{R. Menezes}
\affiliation{Departamento de F\' \i sica, Universidade Federal
da Para\'\i ba, 58051-970 Jo\~ao Pessoa, Para\'\i ba, Brasil.}
\author{C. Wotzasek}
\affiliation{Instituto de F\'\i sica, Universidade Federal do Rio de
Janeiro, 21945-970, Rio de Janeiro, Brazil}

\begin{abstract}
We study duality transformation and duality symmetry in the the electromagnetic-like charged p-form theories.
It is shown that the dichotomic characterization of duality groups as $Z_2$ or $SO(2)$ remains as the only possibilities but are now present in all dimensions even and odd.
This is a property defined in the symplectic sector of the theory both for massive and massless tensors.
It is shown that the duality groups depend, in general, both on the ranks of the fields and on the dimension of the spacetime.
We search for the physical origin of this
two-fold property and show that it is traceable to the dimensional and rank
dependence of the parity of certain operator (a generalized-curl)
that naturally decomposes the symplectic sector
of the action.  These operators are only slightly different in the massive and in
the massless cases but their physical origin are quite distinct.
\end{abstract}
\maketitle


\section{Introduction}
This paper is devoted to the study of
duality groups in theories of complex totally antisymmetric tensors of
arbitrary ranks, both for massless and massive fields, in a
generalized sense which allows to include the proper couplings to external electric and magnetic branes of appropriate dimensions \cite{spa}.
It is our purpose to draw attention to rank and dimensionality dependence of the groups of duality for electromagnetic--like complex $p$-form theories.
This issue is not only of fundamental theoretical importance but may also be relevant for the experimental searches in contexts where magnetic charges play an important role.
In this sense it becomes quite important to study the groups under which duality is manifest.

A given physical phenomenon may, sometimes, possess two equivalent
mathematical descriptions in terms of different sets of variables.
The operation connecting one set into the other is called duality
and maps different regions of the parameter space in a nontrivial way: a model in a
strong coupling regime is related to its dual version working in a weak
coupling regime. This provides valuable information in the study of
strongly interacting models.
Duality has been the subject of intense scrutiny in recent times. Although this subject still lacks complete understanding, there exists now convincing evidence that the focus has been more on its structure and consequences than on testing duality itself.

Duality transformation has had strong impact over different areas of Physics -- from Strings to Condensed Matter \cite{reviews}
Exploring the symmetry under this transformation has been of fundamental importance for investigations in arenas as distinct as quantum field theory, statistical mechanics and string theory.
It is always worth to recall the importance of duality symmetry and the use of totally anti-symmetric tensors in the construction of modern gauge theories.
Antisymmetric tensor  are the building blocks in the construction of gauge theories of extended  objects (strings, membranes,...) \cite{ast}.
The study of $p$-form dualities is becoming more and more important due to recent developments in string theory, where it was shown that inequivalent vacua are related by dualities based on the existence of extended objects, the D-branes \cite{anais}.
In effective field theories they appear naturally as the low-energy dynamics of strings where they
play an important role in the realization of various dualities
among different theories \cite{pol}.

In the last decades we have seem an increasingly greater interest in the theory of (two and more)-dimensional extended objects (membranes and $p$-branes) as unified theories containing non-abelian excitations  \cite{AA} (see also \cite{KY}).
In the electromagnetic scenario, which is the main paradigm for this phenomenon, duality is known to exchange the magnetic with the electric sectors.
Starting with the minimal coupling of the  $p$-form potential $A_p$ to an external source of strength $e$ (usually called as $(p-1)$-brane), duality may then be used to unambiguously define the {\it magnetic} coupling of the extended sources to the dual potential, $\tilde A_q$. Therefore, while the $p$-form potential feels the source as electric, the dual potential $\tilde A_q$ experiences it as magnetic.
It should be observed however that the exchange of the magnetic and electric sectors and the potential and its dual produced by the duality transformation does not, necessarily, occurs as a symmetry for the equations of motion of the system. To display duality as a symmetry, conditions on the dimensionality of the sources or alternatively on the ranks of the potentials become necessary.

Issues of symmetry are in general quite involved and the complete group of duality transformation depends crucially on the field contents of the theory.
In the usual real--potential case, in order to have symmetry under the duality transformation, the dimensions of the electric and magnetic {\it sources} will have to coincide which also imposes equality between the ranks of the potential and its dual. This severely restricts the dimensions where duality may exists as a symmetry and is different for the massive and the massless cases.
In the stage set by electromagnetic theories, such studies have been undertaken both for the massless \cite{massless1, Schwarz:1993vs, massless2} and for the massive cases \cite{DJT,TPvN,noronha} where the dynamical contents are controlled
by Lagrangian densities written in terms of real potential forms, related by duality,
either $A_p$ or $\tilde A_q$, where rank and spacetime dimensions are
related as 
\be
\label{dim1}
D= \begin{cases} p+q+2,&\mbox{for massless tensors}, \\
    p+q+1,&\mbox{for massive tensors}. \end{cases}
\ee

Motivated by the work \cite{BZ}, the analysis of a charged--potential description of the theory becomes necessary which modify the above restrictions, in terms of the dimensions of the sources and ranks of the potentials due to the necessity to introduce a complex field potential.  In this work we study the groups of duality for complex $p$-form potentials $\Psi_{pq} = A_p + i B_q$
or as a real doublet $\left(A_p , B_q\right)$ whose components maintain a relation given by (\ref{dim1}).
Duality is therefore an operation that acts {\it simultaneously} on both fields: $A_p \to \tilde A_q$ and $B_q \to \tilde B_p$ so that $\Psi_{pq}\to \tilde\Psi_{qp}=\tilde A_q + i \tilde B_p$.
In this generalized scenario, not only the proper dynamics are manifest even when fields are electrically and magnetically coupled to sources, but also a complete description of the symmetry groups involved and
their dependence on rank and spacetime dimensions becomes possible of simple analysis.

The real-field situation is then considered as a special case of the complex-tensor field theory which only occur in some special dimensions.
For such dimensions, rank and dimensionality are uniquely connected, implying that one may classify the duality group as dependent on either notions.
It is usual to find a classification of real--tensor duality groups in the literature displaying dimensional dependence.
These results, up to $D=9$, are summarized in the
following table displaying the dimensional dependence of the group
of symmetries for massless and massive $p$-form theories:
\begin{center}
\begin{tabular}{|c|c|c|c|c|c|}
\hline
Dimensions  & massive &  massless  &  rank  &  E/M Brane   \\ \hline
9  &  SO(2)   &         &  4  &   3-brane \\ \hline
8  &          &  SO(2)  &  3  &   Membrane  \\ \hline
7  &  $Z_{2}$   &         &  3  &    Membrane  \\ \hline
6  &          &  $Z_{2}$  &  2  &    String   \\ \hline
5  &  SO(2)   &         &  2  &    String  \\ \hline
4  &          &  SO(2)  &  1  &    Monopole   \\ \hline
3  &  $Z_{2}$   &         &  1    & Monopole   \\ \hline
2  &          &  $Z_{2}$  &  0  &    Instanton  \\ \hline
\end{tabular}
\end{center}
This table displays the characteristics of all possible self-dual and SO(2) symmetric actions described by a real potentials and the electric/magnetic branes to which they are coupled.
It shows that massless duality symmetry is restricted to even dimensional spacetimes while massive duality symmetry happens exclusively in odd dimensional spacetimes.  Still self-duality is not present in all cases since it only happens for half of the cases in both situations.
For massless and massive fields it happens in dimensions $D=2\,\mbox{mod(4)}$ and $D=3\,\mbox{mod(4)}$, respectively, where the group structure is $Z_2$.  In the other situations the group is $SO(2)$ and there is no self-duality.

The continuous one-parameter $SO(2)$ symmetry means that the electric and the magnetic sectors rotate into each other.  This is the usual meaning of duality or duality rotation from textbook electromagnetism.
The $Z_2$ symmetry on the other hand means that the action diagonalizes into self-dual and anti self-dual sectors, each carrying half the number of degrees of freedom.  If rank and dimension are such that the theory displays the $Z_2$ symmetry then the individual, self and anti self-dual components may be obtained alternatively through the imposition of constraints of self-duality.  

For the more general situations, to be developed here, the structure of the duality group is naively expected to be dependent both on the dimensionality and on the rank of the potential.
Our study will clearly show the existence of a dichotomy associated to the groups of duality as consequence of the only two possible ways to decompose the symplectic sector of the complex--field theory.
Our analysis shows that the physical property triggering the proper symplectic sector or group of duality is the parity of a certain generalized curl. The parity dependence on rank and dimension is rather simple in the real--potential case but the analysis of the complex--field case is quite involved. Details of this analysis, both for massless and massive tensors, are shown in the next Section.

\section{Complex-potential duality}

This section is devoted to our main result -- to study issues of duality for a
complex-field theory, both massless and massive.
In particular we are interested in the analysis of the
rank and dimensional dependence of the groups of duality.
We start this section
with the massless case and complete the work in the following
subsection with the analysis of the massive case.
The question that poses itself immediately is about the possible presence of other groups of duality as a consequence of the generalization of the theory.  This question will be clearly answered by the dual projection approach after the factorization of the symplectic sector of the theory.

\subsection{The massless case}

Let us consider the massless Lagrangian density, having complementary tensorial indices, defined by a the real part of field tensor squared as usual, leading to the following action
\begin{equation}
\label{010} {\cal S} = \left\langle
\frac{(-1)^p}{2(p+1)!}
[F_{p+1}\left(A_p\right)]^2  +
\frac{(-1)^q}{2(q+1)!}
[F_{q+1}\left(B_q\right)]^2 \right\rangle\; .
\end{equation}
Here the field strength reads
$F_{p+1}\left(A_p\right)= F_{\mu_1 \mu_2 \ldots \mu_{p+1}}=\partial_{[\mu_1}A_{\mu_2\cdots\mu_{p+1}]}$ and similarly for $F_{q+1}\left(B_q\right)$ and $\langle\cdots\rangle$ denotes integration over the $D$ dimensional manifold. The choice of the coefficient adopted here is justified in order to normalize the symplectic sector in the usual way. We use the following metric
$g^{\mu\nu}=diag(+,-,\cdots ,-)$. 

For massless cases, the first-order reduction is done non-covariantly,
leading to
\ba \label{020} &{\cal S}& =\Big\langle\bar\pi_p \dot
{\bar A}_p -  \fr{(-1)^p}{2p!}\bar\pi_p^2
           + \bar A_{0,p-1}\left(\nabla \cdot {\bar \pi}_p\right) -\fr{(-1)^p}{2p!} \bar{\cal B}_p^2 \nn
          &+& \bar\tau_q \dot{\bar B}_q -  \fr{(-1)^q}{2q!} \bar\tau_p^2 +\bar B_{0,q-1}\left(\nabla \cdot
          \bar\tau_q\right) -   \fr{(-1)^q}{2q!} \bar{\cal B}_q^2 \Big\rangle
\ea where $\bar\pi_p$ and $\bar\tau_q$ are $p$ and $q$-forms
playing the role of canonical momenta for the original fields
$\bar A_p$ and $\bar B_q$ respectively, taking values over the $D-1$ dimensional Euclidean spatial
manifold. 
We keep in mind that $p$ e $q$ have
complementary tensorial ranks, with $p+q=D-2$, where $D$ is the
Minkowski spacetime dimension which is characteristic of massless duality.  The role of magnetic fields for
the $q$ and $p$-form potentials $\bar A_p$ and $\bar B_q$, are played
by $\bar{\cal B}_q = (\widehat{\epsilon\partial} A_p)$ and
$\bar{\cal B}_p = (\widehat{\epsilon\partial} B_q)$, respectively, where the (spatial)
curl-operator is,
\begin{equation} \label{a30} (\widehat{\epsilon\partial})
\equiv \epsilon_{k_1 k_2 \cdots k_{D-1}}\partial_{k_{D-1}}
\end{equation}
with $k_i=1,2,\cdots,D-1$ taking values on the spatial indices
only.
Here $\bar A_{0,p-1}=\bar A_{0i_1 i_2 \ldots i_{p-1}}$ and $\bar
B_{0,q-1}=\bar B_{0i_1 i_2 \ldots i_{q-1}}$ are Lagrange
multiplier fields imposing the extended Gauss law, \ba \label{}
\nabla \cdot \bar\pi_p \approx 0\nn \nabla \cdot \bar\tau_q\approx
0 \ea as first-class constraints.  The solution of these
constraints is given in terms of a curl-operator
 ($\widehat{\epsilon\partial}$) and some new $q$ and $p$-forms,
 ${{\bar B^*}}_q$ and $ {{\bar A^*}}_p$, as
\ba \bar\pi_p &=& \widehat{\epsilon \partial} {{\bar B^*}}_q \nn
\bar\tau_q &=& \widehat{\epsilon \partial} {{\bar A^*}}_p \ea This
is a crucial step leading from the phase-space of a single complex-field
theory to the configuration space of two complex-field theory, albeit
in first-order. The next step is the redefinition of the above
space producing a more symmetric presentation for the theory as
\ba
\bar A_p  &=& \bar A_p^+ + \bar A_p^- \nn {{\bar A^*}}_p
           &=&\eta \left(\bar A_p^+ - \bar A_p^- \right)
\ea
and
\ba
\bar B_q
&=&\bar B_q^+ +\bar B_q^- \nn {{\bar B^*}}_q  &=&\eta \left(\bar
B_q^+ -\bar B_q^- \right) \; ,
\ea
where $\eta=\pm 1$ is the signature of the transformation.
Upon these redefinitions, the
action (\ref{020}) is rewritten in terms of the new fields as
\ba
\label{021} {\cal S} &=&\Big\langle \eta \Big(-\bar A^+_p
\widehat{\epsilon \partial} \dot{\bar B}_q^+  -\bar B^+_q
\widehat{\epsilon \partial} \dot{\bar A}_p^+
        + \bar A^-_p \widehat{\epsilon \partial} \dot{\bar B}_q^-
        + \bar B^-_q \widehat{\epsilon \partial} \dot{\bar A}_p^-  \nn
     && + \bar A^+_p \widehat{\epsilon \partial} \dot{\bar B}_q^-
        - \bar B^-_q \widehat{\epsilon \partial} \dot{\bar A}_p^+
        - \bar A^-_p \widehat{\epsilon \partial} \dot{\bar B}_q^+
        + \bar B^+_q \widehat{\epsilon \partial} \dot{\bar A}_p^- \Big) \nn
     && -  \fr{(-1)^q}{q!} \left[\left(\epsilon \partial A^{(+)}_p \right)^2
        + \left(\epsilon \partial A^{(-)}_p \right)^2\right] \nn
     &&-  \fr{(-1)^p}{p!} \left[\left(\epsilon \partial B^{(+)}_q \right)^2
        + \left(\epsilon \partial B^{(-)}_q \right)^2\right] \Big\rangle
\ea
Inspection shows the presence, in the symplectic sector, after
some integrations by parts, of four types of terms obtained from
the combinations of $\bar A_p^\pm$ and $\bar B_q^\pm$ and the curl
operator $(\widehat{\epsilon \partial})$.  
The presence of all these terms simultaneously in the action is
illusory since some terms, depending on the parity of the operator,
become total
derivatives.  Since the parity property  seems to be strongly
dependent on rank and dimensionality, it is important to carefully analyze this dependence.
This goes as follows. The parity of the
generalized-curl is defined as
 \ba
<\bar A_p \widehat{\epsilon \partial}  \bar B_q>
= \hat{\cal P}{(p ,q)} <\bar B_q
\widehat{\epsilon \partial}  \bar A_p> \ea
and its dependence on
the rank of the $p$ and $q$-forms or dimensionality is given as
\ba
\label{parity} \hat{\cal P}{(p ,q)} &=&  (-1)^{(p+1)(q+1)}
\nn &=& (-1)^{(D+1)(q+1)} = (-1)^{(D+1)(p+1)}
\ea
Using this relation, the theory (\ref{021}) gets reorganized as
\ba
\label{022} {\cal S} &&= \Big \langle \eta \Big(- \Big[ 1- \hat{\cal P}{(p ,q)} \Big]\, \bar B^+_q
\widehat{\epsilon \partial}\dot{\bar A}_p^+ \nn
            &&  + \Big[ 1- \hat{\cal P}{(p ,q)} \Big]\, \bar B^-_q \widehat{\epsilon \partial} \dot{\bar A}_p^-  \nn
           && - \Big[ 1+ \hat{\cal P}{(p ,q)} \Big]\, \bar B^-_q \widehat{\epsilon \partial} \dot{\bar A}_p^+
              + \Big[ 1+ \hat{\cal P}{(p ,q)} \Big]\, \bar B^+_q \widehat{\epsilon \partial} \dot{\bar A}_p^- \Big) \nn
               && -  \fr{(-1)^q}{q!} \left[\left(\epsilon \partial A^{(+)}_p \right)^2
        + \left(\epsilon \partial A^{(-)}_p \right)^2\right]\nn
      &&-  \fr{(-1)^p}{p!} \left[\left(\epsilon \partial B^{(+)}_q \right)^2
        + \left(\epsilon \partial B^{(-)}_q \right)^2\right] \Big \rangle
\ea
displaying the presence of a $Z_2$ or
$SO(2)$ symmetry depending upon the parity of the symplectic
operator, as expected from our previous discussions.   For the odd
 parity instances, the action displays a clear $Z_2$ symmetry while the parity
  even instances displays a $SO(2)$ symmetry. Therefore, there are only two possibilities and they clearly result from the parity of the operator solving the Gauss law.

Next, let us firstly verify the real-field limit before discussing the consequences of our findings.
The usual self-duality where $p=q$ and $\bar A= \bar B$ is easily
recovered but (\ref{022}) displays more general possibilities. For
the massless case, the minimal requirement to self-duality is
$D=p+q+2 \to 2(p+1)$, i.e., real-field self-duality is restricted
to even dimensional spacetimes. Thereupon we clearly conclude
that if: (i) $D=4k$, the first line the (\ref{022}) vanishes and
we obtain a SO(2) structure, while (ii) $D=4k+2$, only the first
line in (\ref{022}) survives, resulting in the $Z_2$ structure
mentioned above. Therefore action (\ref{022}) reproduces the well
known structure for the real-field self-duality \cite{massless1,massless2}.  Next we
analyze the new predictions of this action.

For $p \neq q$, the specific group
structure will depend on rank and dimensionality in a rather
involved way. For (i) odd dimensional spacetimes, a simple
analysis of the parity of the curl operator defined in
(\ref{parity}) shows
\begin{equation} \label{023}
\hat{\cal P}{(p ,q)} \xrightarrow{D=2k+1} +1
\end{equation}
 leading to an action as
\ba
 \label{024} {\cal S} =&&  \Big\langle - 2 \eta
\Big[  \bar B^-_q \widehat{\epsilon \partial} \dot{\bar A}_p^+ - \bar B^+_q \epsilon \partial \dot{\bar A}_p^- \Big]
\nn &-&\fr{(-1)^q}{q!} \Big[\left(\epsilon \partial A^{(+)}_p \right)^2
        + \left(\epsilon \partial A^{(-)}_p \right)^2\Big] \nn
      &-&  \fr{(-1)^p}{p!} \left[\left(\epsilon \partial B^{(+)}_q \right)^2
        + \left(\epsilon \partial B^{(-)}_q \right)^2\right]  \Big\rangle
\ea
that can be reordered to display an explicitly $SO(2)$ symmetry
\ba
\label{0024}
{\cal S}=\Big\langle  2 \eta \Big[  \bar B^\alpha_q
\epsilon^{\alpha\beta}\widehat{\epsilon \partial} \dot{\bar
A}_p^\beta\Big]&-&\fr{(-1)^q}{q!} \left(\epsilon \partial A^{(\beta)}_p \right)^2
     \nn &-&  \fr{(-1)^p}{p!} \left(\epsilon \partial B^{(\beta)}_q \right)^2
    \Big\rangle
\ea
where $\epsilon^{+-}=1$. This equation is invariant under the action of the following dual transformation acting on the double internal space
\ba \label{025}
\begin{pmatrix} \bar
A_p^+ \\
    \bar A_p^-
\end{pmatrix}
     \to
\begin{pmatrix}
     \cos\theta&\sin\theta \\
    -\sin\theta&\cos\theta
\end{pmatrix}
      \begin{pmatrix} \bar A_p^+\\
    \bar A_p^-\end{pmatrix}
\ea
\ba \label{02521}
\begin{pmatrix} \bar
B_p^+ \\
    \bar B_p^-
\end{pmatrix}
     \to
\begin{pmatrix}
     \cos\theta&\sin\theta \\
    -\sin\theta&\cos\theta
\end{pmatrix}
      \begin{pmatrix} \bar B_p^+\\
    \bar B_p^-\end{pmatrix}
\ea
where $\theta$ is the $SO(2)$ group parameter.  For the (ii) even dimensional case, the linking between rank and dimensionality demands the same parity for the $p$ and $q$ ranks leading to a two-fold consequence, depending if the pair ($p,q$) is even or odd.
\begin{equation}
\label{026}
\hat{\cal P}{(p ,q)} \xrightarrow{D=2k}
    \begin{cases} -1,&\mbox{if $(p,q)$ even}, \\
    +1,&\mbox{if $(p,q)$ odd}, \end{cases}
\end{equation}
For the odd rank case we, once again, have even parity for the curl operator, leading to the $SO(2)$ Lagrangian (\ref{0024}) as before. For the even rank case, on the other hand, the parity for the curl
 operator is even, leading instead to $Z_2$ symmetry
  ($\bar A_p^\pm(\bar B_q^\pm)\rightleftharpoons \bar A_p^\mp(\bar B_q^\mp)$), controlled by the following action
\ba
\label{027}
{\cal S} =&&\Big\langle - 2 \eta \Big[  \bar B^+_q \widehat{\epsilon \partial} \dot{\bar A}_p^+
              -  \bar B^-_q \widehat{\epsilon \partial} \dot{\bar A}_p^-\Big]
         \nn&& -\fr{(-1)^q}{q!} \left[\left(\epsilon \partial A^{(+)}_p \right)^2
        + \left(\epsilon \partial A^{(-)}_p \right)^2\right] \nn
          &&-\fr{(-1)^p}{p!} \left[\left(\epsilon \partial B^{(+)}_q \right)^2
        + \left(\epsilon \partial B^{(-)}_q \right)^2\right]  \Big\rangle
\ea
or
\begin{equation}
{\cal S} = {\cal S}^{+}(A_p^{+},B_q^{+}) + {\cal S}^{-}(A_p^{-},B_q^{-})
\end{equation}
where
\ba
{\cal S}^{\pm}(A_p^{\pm},B_q^{\pm})= \Big\langle \mp 2 \eta  \bar B^\pm_q \widehat{\epsilon \partial} \dot{\bar A}_p^\pm
                      &&-\fr{(-1)^q}{q!}\left(\epsilon \partial A^{(\pm)}_p \right)^2 \nn
                      && -  \fr{(-1)^p}{p!}\left(\epsilon \partial B^{(\pm)}_q \right)^2
         \Big\rangle \; .\nonumber
\ea
This result extends and generalizes the real-field theory
and concludes our analysis of the massless complex-field duality group.  Next we use the dual projection approach to disclose the group of duality structure for Proca-like complex-field theories.

\subsection{The massive case}

Let us consider the massive case, defined by a complex--field Lagrangian 
\ba
{\cal S} &=& \Bigg\langle \frac{(-1)^p \,q!}{2(p+1)!}  F_{p+1}\left(A_p\right)F^{p+1}\left(A^p\right) -
 (-1)^{p} \frac{m^2}{2} A_p A^p \nn
                     &&+  \frac{(-1)^q \,p!}{2(q+1)!} H_{q+1}\left(B_q\right)H^{q+1}\left(B^q\right) -
 (-1)^{q} \frac{m^2}{2} B_q B^q \Bigg\rangle \nonumber
\ea
Here, distinctly from the massless case, the p-form fields take values over the $D$ dimensional
Minkowski spacetime manifold.
Notice that due to the absence of a Gauss law, the first-order reduction can
 be effected with the introduction of two auxiliary fields, in a covariant fashion, as
\ba
\label{045}
{\cal S} &=&  \Bigg \langle \Pi_q \epsilon \partial A_p  - \frac{(-1)^q} 2 \Pi^2_q - (-1)^{p} \fr{m^2}{2}  A_p A^p \nn
          &&+ \wt{\Pi}_p \epsilon \partial B_q  - \frac{(-1)^p} 2 \wt\Pi^2_p - (-1)^{q} \fr{m^2}{2}  B_q B^q \Bigg \rangle
\ea
We shall denote the first term in each line of this
action as the ``covariant symplectic sector" and the remaining of
the action as the symplectic potential or just potential for short
\cite{fadjack}. This terminology allows us to follow a quick parallel with the massless case.
Here $\Pi_q$ is a $q$-form auxiliary field and our
notation is
\begin{equation}
\Pi_q\;\epsilon \partial A_p = \epsilon^{\mu_1\cdots\mu_q\alpha\mu_1\cdots\mu_p}
\Pi_{\mu_1\cdots\mu_q}\partial_\alpha A_{\nu_1\cdots\nu_p}\, .
\end{equation}
This brings naturally the generalized-curl operator
($\epsilon\partial$) into the ({\it covariant}) symplectic sector
of the action but it clearly has a different origin than the ($\widehat{\epsilon\partial}$) operator. For the massless $p$-forms potentials the ({\it non-covariant}) curl operator was brought into the
symplectic part of the action as the consequence of the solution
of the Gauss law. There is no gauge symmetry here.
Other notations and conventions follow those of the preceding subsection but we have changed the normalizations to keep the (covariant) symplectic sector with the usual structure.
The next step is the redefinition of the above space producing a more symmetric presentation
for the theory as
\ba
A_p       &=& A_p^+ + A_p^- \nn
\Pi_p  &=& \eta m \left(A_p^+ - A_p^-\right)
\ea
and
\ba
B_q       &=& B_q^+ + B_q^- \nn
\wt\Pi_q            &=& \eta m \left( B_q^+ - B_q^- \right) \; .
\ea
which disclose the internal space of potentials, now in covariant form, as it is adequate for massive forms.
Upon these redefinitions, the action (\ref{045}) is rewritten in terms of the new fields as
\ba
\label{055}
{\cal S} = &&  \Bigg \langle \eta m \Big( A^+_p \epsilon \partial B_q^+       + B^+_q \epsilon \partial {A}_p^+
             - A^-_p \epsilon \partial {B}_q^-     - B^-_q \epsilon \partial {A}_p^-  \nn
          && + A^+_p \epsilon \partial {B}_q^- - B^-_q \epsilon \partial {A}_p^+
             - A^-_p \epsilon \partial {B}_q^+ + B^+_q \epsilon \partial {A}_p^-  \Big)\nn
           &&  + (-1)^{p+1} \fr {m^2}2 \left[(A_p^{+})^2+(A_p^{-})^2\right] \nn &&+ (-1)^{q+1}
           \fr {m^2}2 \left[(B_q^{+})^2+(B_q^{-})^2\right]  \Bigg \rangle
\ea
The classifications of the duality groups, here too, will be a consequence of the dependence on
rank and dimensionality of the parity property of the covariant curl-operator $\epsilon\partial$, defined as
\ba
<a_p \epsilon \partial  B_q>
= {\cal P}{(p ,q)} <B_q \,\epsilon \partial  A_p>
\ea
whose dependence on the rank of the $p$ and $q$-forms or dimensionality and rank is  given as
\ba \label{covparity}
{\cal P}{(p ,q)} &=&  (-1)^{(p+1)(q+1)}\nn
&=& (-1)^{D(q+1)} = (-1)^{D(p+1)}
\ea
leading to the following action
\ba
\label{01010}
{\cal S} &=& \Big\langle \eta m \Big( [ 1+ {\cal P}{(p ,q)}]\, B^+_q \epsilon \partial {A}_p^+
              - [ 1+ {\cal P}{(p ,q)} ]\, B^-_q \epsilon \partial {A}_p^-  \nn
           && - [ 1- {\cal P}{(p ,q)} ]\, B^-_q \epsilon \partial {A}_p^+
              + [ 1- {\cal P}{(p ,q)} ]\, B^+_q \epsilon \partial {A}_p^- \Big) \nn
           &&                + (-1)^{p+1} \fr {m^2}2 \left[(A_p^{+})^2+(A_p^{-})^2\right]\nn &&+ (-1)^{q+1}
           \fr {m^2}2 \left[(B_q^{+})^2+(B_q^{-})^2\right]             \Big\rangle.
\ea

The usual real-field duality symmetry where $p=q$ and $A=B$ is easily retrieved. For the massive case,
 the minimal requirement is $D=p+q+1 \to 2p+1$, i.e., real-field duality symmetry
is restricted to odd dimensional spacetimes. Clearly we conclude that if: (i) $D=4k+1$, the
first line the (\ref{01010}) vanishes and we obtain a SO(2) structure, while (ii) $D=4k-1$,
only the first line in (\ref{01010}) survives, resulting in the $Z_2$ structure mentioned above.
Therefore action (\ref{01010}) reproduces the well known structure for real-field duality symmetry \cite{TPvN,noronha}.

For  $p \neq q$, the specific group structure will again depend on rank and dimensionality
 in a rather involved way.  For (i) even dimensional spacetimes, a simple analysis of the parity of the
  covariant curl-operator defined in (\ref{covparity}) shows
\begin{equation}
\label{023a}
{\cal P}{(p ,q)}  \xrightarrow{D=2k} + 1
\end{equation}
leading to an explicitly $Z_2$ action as
\ba
\label{024a}
{\cal S} &=& \Bigg\langle 2 \eta m \Big[ B^+_q \epsilon \partial {A}_p^+
              -  B^-_q \epsilon \partial {A}_p^- \Big] \nn
 &+& (-1)^{p+1} \fr {m^2}2 \left[(A_p^{+})^2+(A_p^{-})^2\right]\nn &+& (-1)^{q+1}
           \fr {m^2}2 \left[(B_q^{+})^2+(B_q^{-})^2\right]
              \Bigg\rangle
\ea
or
\begin{equation} \label{024b}
{\cal S} = {\cal S}^{+}(A^+_p,B_q^+) + {\cal S}^{-}(A^-_p,B_q^-)
\end{equation}
where
\ba
{\cal S}^\pm(A^\pm_p,B_q^\pm) = \Bigg\langle \pm 2 \eta m \Big[ B^\pm_q \epsilon \partial {A}_p^\pm \Big]
 + (-1)^{p+1} \fr {m^2}2 (A_p^{\pm})^2 \nn + (-1)^{q+1}
           \fr {m^2}2 (B_q^{\pm})^2
              \Bigg\rangle \nonumber
\ea
invariant under the action ($A_p^\pm(B_q^\pm)\rightleftharpoons A_p^\mp(B_q^\mp)$) the  dual transformation
For the (ii) odd dimensional case, the linking between rank and dimensionality demands the same parity for the $p$ and $q$ ranks leading to a two-fold consequence, depending if the pair ($p,q$) has even or odd parity as,
\begin{equation}
\label{026a}
{\cal P}{(p ,q)}  \xrightarrow{D=2k+1}
\begin{cases}-1,&\mbox{if $(p,q)$ even},\\
    +1,&\mbox{if $(p,q)$ odd}, \end{cases}
\end{equation}
For the odd rank case we, once again, have even parity for the curl operator, leading to $Z_2$ obtaining
the Lagrangian (\ref{024b}) as before. For the even rank case we, the parity for the
curl-operator is odd, leading the following action
\ba
\label{027a}
{\cal S} &=&\Bigg\langle 2\eta m \Big[ B^+_q \epsilon \partial {A}_p^- -
              B^-_q \epsilon \partial {A}_p^+ \Big]\nn
                &+& (-1)^{p+1} \fr {m^2}2 \left[(A_p^{+})^2+(A_p^{-})^2\right] \nn
               &+& (-1)^{q+1}  \fr {m^2}2 \left[(B_q^{+})^2+(B_q^{-})^2\right]
              \Bigg\rangle,
\ea
that can be rewritten as
\ba
\label{028j}
{\cal S} = \Bigg\langle 2\eta m B^{\alpha}_q \epsilon^{\alpha \beta} \epsilon \partial {A}_p^{\beta}
                &+& (-1)^{p+1} \fr {m^2}2 \left(A^{\beta}_p\right)^2 \nn
               &+& (-1)^{q+1}  \fr {m^2}2 \left(B^{\beta}_q\right)^2
              \Bigg\rangle,
\ea
displaying instead the explicit $SO(2)$ symmetry. The analysis just effected show how the duality symmetry, either
$SO(2)$ or $Z_2$, restricted to odd-dimensional spacetimes in the massive real-field case, gets extended to all dimensions, both odd and even, in the more general context of complex-fields studied here.

\section{Conclusions}

We have provided a classification for the duality groups for theories containing complex-fields whose components are two complementary form fields in spacetimes of arbitrary dimensions.
Our results are as follows. 
We provided a clear cut proof that there are only two possible groups of dualities, either $Z_2$ or $SO(2)$ that decompose the symplectic sector of the theory but Hamiltonian is not sensitive of these properties.  Differently from the real-field case, the groups may depend on rank and dimensionality of the spacetime.  
We provided a new definition for self-duality that extends the usual case of real-tensors and reduces to it in the special cases. It is important to stress that both massive or massless tensors, it is the parity dependence on rank and dimensionality of the curl-operators that decides the group structure. The technique that is used to establish these results is called dual projection. It introduces an internal space of potentials where duality is naturally defined. It should be noticed that, although these operators are only slightly different in the massive and in
the massless cases, their physical origin are quite distinct. However the final outcome is basically the same in the sense that they allow us to separate the symplectic sector in two distinct pieces. The dual symmetric actions displaying all dualities manifestly are construct for all cases using the dual projection.

The physical results obtained, after elaboration upon these analysis, are displayed in the following tables.  For massless $p$-form and $q$-forms potentials,
the symmetry groups are as follows,
\vspace{1cm}
\begin{center}
TABLE 2 - MASSLESS
\end{center}
\begin{center}
\begin{tabular}{|c|c|c|c|}
\hline
\,\,Dimension\,\, & \,\,p rank & q rank\,\, & \,\,Symmetry\,\, \\
\hline
2  & 0 & 0 & $Z_2$ \\
\hline\hline
3  & 0 & 1 & $SO(2)$ \\
\hline\hline
4 & 0 & 2 & $Z_2$ \\
 \cline{2-4}
 & 1 & 1 & $SO(2)$ \\
\hline\hline
5 & 0 & 3 & $SO(2)$ \\
\cline{2-4}
 & 1 & 2 & $SO(2)$ \\
\hline\hline
 & 0 & 4 & $Z_2$ \\
 \cline{2-4}
6 & 1 & 3 & $SO(2)$ \\
\cline{2-4}
 & 2 & 2 & $Z_2$ \\
\hline\hline
& 0 & 5 & $SO(2)$ \\
 \cline{2-4}
7 & 1 & 4 & $SO(2)$ \\
\cline{2-4}
 & 2 & 3 & $SO(2)$ \\
 \hline
\end{tabular}
\end{center}
For odd dimensions the group is always $SO(2)$ but the ranks are of opposite parity. For even dimensions both structures are manifest but ranks must have the same parity either even or odd.  In even dimensions the group is $Z_2$ for even (p,q) ranks and again $SO(2)$ for odd (p,q) ranks.

For massive forms the group dependence is somehow complementary and is displayed by the following table,
\vspace{.1cm}
\begin{center}
TABLE 3 - MASSIVE
\end{center}
\begin{center}
\begin{tabular}{|c|c|c|c|}
\hline
\,\,Dimension\,\, & \,\,p rank\,\, & \,\,q rank\,\, & \,\,Symmetry\,\, \\
\hline\hline
2  & 0 & 1 & $Z_2$ \\
\hline\hline
3 & 0 & 2 & $SO(2)$ \\
\cline{2-4}
 & 1 & 1 & $Z_2$ \\
\hline\hline
4 & 0 & 3 & $Z_2$ \\
\cline{2-4}
 & 1 & 2 & $Z_2$ \\
\hline\hline
 & 0 & 4 & $SO(2)$ \\
 \cline{2-4}
5 & 1 & 3 & $Z_2$ \\
\cline{2-4}
 & 2 & 2 & $SO(2)$ \\
\hline\hline
 & 0 & 5 & $Z_2$ \\
 \cline{2-4}
6 & 1 & 4 & $Z_2$ \\
\cline{2-4}
 & 2 & 3 & $Z_2$ \\

\hline
\end{tabular}
\end{center}
For even dimensions the group structure is unique equals $Z_2$ but again $p$ and $q$ must be of opposite parity. On the other hand, both $Z_2$ and $SO(2)$ are manifest in odd dimensions. It is, however, $Z_2$ for odd rank tensors (p,q) and $SO(2)$ for even (p,q) ranks.
We believe that this study completely characterizes the duality group for complex-tensor electromagnetic like theories, both massless and massive, for all dimensions and all ranks.

{\bf Acknowledgments}  This work is partially supported by CNPq, FUJB, and CAPES,
Brazilian scientific agencies and by PRONEX. RM thanks the Physics Department of UFRJ for the kind hospitality
during the course of the investigation and the PROCAD program for financial support.


\ed


To The Editor,
Physics Letter B
Professor P.V. Landshoff,
Dept. of Applied Mathematics and Theoretical Physics,
University of Cambridge, Wilberforce Road, Cambridge CB3 9EW, UK.
p.v.landshoff@damtp.cam.ac.uk

Dear Editor,

we are submitting our paper entitled "ON DUALITY SYMMETRY IN CHARGED P-FORM THEORIES" to be considered for
publication in the Physics Letter B.

                              Thanking you,
                            Yours sincerely,
                              C.Wotzasek
                       e-mail: clovis@if.ufrj.br